\documentclass[epsf]{svjour2}                    
\smartqed  
\usepackage{graphicx}
\begin{document}

\title{Markov Properties of Electrical Discharge Current Fluctuations in
Plasma}


\author{S. Kimiagar, M. Sadegh Movahed, S. Khorram  and M. Reza Rahimi Tabar}


\institute{S. Kimiagar \at
              Department of Physics,science faculty, Tehran central branch, Islamic Azad University, Tehran,
Iran\\
              \email{kimia@khayam.ut.ac.ir}           
           \and
           M. Sadegh Movahed \at
              Department of Physics, Shahid Beheshti university, G.C., Evin, Tehran 19839, Iran\\
School of Astronomy, Institute for Research in Fundamental Sciences, (IPM), P. O. Box 19395-5531, Tehran, Iran\\
              \email{m.s.movahed@ipm.ir}\and
              S. Khorram \at
              Research Institute for Applied Physics and Astronomy, University of Tabriz, Tabriz 51664, Iran\\
              \email{sirousk@yahoo.com} \and
              M. Reza Rahimi Tabar \at
              Department of Physics, Sharif University of
Technology, P.O.Box 11365-9161, Tehran, Iran\\
Fachbereich Physik, Universit\"{a}t Osnabr\"{u}ck, Barbarastra{\ss}e 7, 49076 Osnabr\"{u}ck, Germany\\
\email{mohammed.r.rahimi.tabar@uni-oldenburg.de} }

\date{Received: date / Accepted: date}

\maketitle

\begin{abstract}
Using the Markovian method, we study the stochastic nature of
electrical discharge current fluctuations in the Helium plasma.
Sinusoidal trends are extracted from the data set by the
Fourier-Detrended Fluctuation analysis and consequently cleaned data
is retrieved. We determine the Markov time scale of the detrended
data set by using likelihood analysis. We also estimate the
Kramers-Moyal's coefficients of the discharge current fluctuations
and derive the corresponding Fokker-Planck equation. In addition,
the obtained Langevin equation enables us to reconstruct discharge
time series with similar statistical properties compared with the
observed in the experiment. We also provide an exact decomposition
of temporal correlation function by using Kramers-Moyal's
coefficients. We show that for the stationary time series, the two
point temporal correlation function has an exponential decaying
behavior with a characteristic correlation time scale. Our results
confirm that, there is no definite relation between correlation and
Markov time scales. However both of them behave as monotonic
increasing function of discharge current intensity. Finally to
complete our analysis, the multifractal behavior of reconstructed
time series using its Keramers-Moyal's coefficients and original
data set are investigated. Extended self similarity analysis
demonstrates that fluctuations in our experimental setup deviates
from Kolmogorov (K41) theory for fully developed turbulence regime.

\end{abstract}

\section{Introduction}

Many natural phenomena are identified by a degree of stochasticity.
Turbulent flows, seismic recordings and plasma fluid are a few
examples of such phenomena
\cite{sadeghcmb1,1,3,6,8,kimiadfa,Jafari03,kantz,tabar06,Ghasmei05,chi05,physa,li07,li06}.
Interpretation and estimation of physical and chemical properties of
plasma fluid have been one of the main research areas in the science
of electromagnetic hydrodynamics.  It is well-known that discharge
current fluctuations in the plasma often exhibits irregular and
complex behavior \cite{li07,li06}. Fluctuations happening commonly
in the plasmas of gas discharges are responsible for a significant
aliquot part of physical and chemical properties of plasma and
reveal many interesting manner of plasma dynamics, energy carrier,
neutrality and shielding areas form microscopic as well as
macroscopic points of view. Generally,  fluctuations  arising in
plasma can be classified as: fluctuations in the electron
temperature, fluctuations in the local plasma space-potential which
is assumed to be periodic and fluctuations in the local random
electrical current in the plasma \cite{1}.  It is supposed that,
these kinds of fluctuations can be characterized according to the
following main properties. The statistics of time series are roughly
homogeneous and isotropic, the spatial correlation length scale is
shorter than the background charge density and temperature flow
characteristic length scales. The autocorrelation function of such
process is also demonstrated as anti-correlated behavior (see e.g.
\cite{kimiadfa,li07,li06}). The phase space of relevant factors
which influence the trajectory of current fluctuations measured by
Langmuir probe in the plasma is enormously large. Therefore, there
is no remedy to use stochastic tools for investigating their
statistical properties. In the presence of complexity as well as
non-linearity in a typical plasma fluctuations, traditional methods
in data analysis encounter with spurious or at least give unreliable
results for explanation of plasma dynamics.

It is impossible to consider all
physical features of plasma fluctuations as a turbulent transport in
the context of deterministic methods. Also finding a good agreement
between properties of plasma discharge fluctuations and that of for
a turbulence regime can play an important role to track the
dissipation of energy transfer at different scale in the plasma
fluid for various values of current intensity by means of
multiplicative cascade model.

There are many stochastic analysis devoted to study the plasma
fluctuations. Fluctuations of electric and magnetic fields of
plasma, spectral density, logistic mapping and nonlinearity of
ionization wave have been investigated in Refs.
\cite{1,3,6,8,kimiadfa,9,10,11}. For example,  Carreras {\it et
al.}, have demonstrated that plasma fluctuations behave as a
multifractal process with nonlinearity comparable to the fluid
turbulence \cite{carr0}. Also Budaev {\it et al.},
\cite{budev05,budev04} by using the scaling behavior of structure
function and wavelet transform modulus maxima (WTMM), showed that
the anomalous transport of particles in the plasma has multifractal
nature. The universality of stochastic properties of different
plasma with various experimental equipments as well as different
physical and chemical operating regimes, have also been explored in
some previous studies. These universality which can be determined in
experiment are led to insights through the understanding of plasma
dynamics. Universality in power spectrum of plasma fluctuations for
various plasma has been investigated in \cite{pedro99}.

Although the analysis of fluctuations in plasma has a long history,
there are, nevertheless, some important issues, such as stochastic
features and most important characteristic time and length scales in
the presence of nonstationarity and trends have remained unexplained
so far. Recently a robust statistical method has been developed to
explore an effective equation that can reproduce stochastic data
with an accuracy comparable to the measured one
\cite{Jafari03,kantz,Sieger,fri97,fri00,fri97b,fri02}. As in many
early researches has been confirmed, one may utilize it to:
\\$1:$ reconstruct the original process with similar statistical
properties, and \\$2:$ understand the nature and properties of the
stochastic process \cite{Jafari03,kantz,tabar06,Ghasmei05,Sieger}.
One of a main task in the evolution of plasma fluctuations is to
quantify most relevant statistical properties such as structure
function, characteristic time and length scales, drift and diffusion
coefficient. In addition, deriving a reliable stochastic model to
reconstruct the data measured by Langmuir probe will be of interest
in the simulation of plasma fluctuation.

Actually, data measured by Langmuir probe integrate fluctuations in
plasma quantities such as number density of charge carriers,
electron temperature, floating potential, heat transport and so on,
consequently the statistical characteristics of probe fluctuations
can be used to obtain information from physical and chemical points
of view. For these purpose we use a robust methods in the complex
systems namely, Markovian method to analyze the plasma fluctuations
during steady state plasma discharge.

Due to the many limitations in an experimental setup for measuring
desired fluctuations, the original fluctuations may be influenced by
some trends and nonstationarities. In addition, the following
necessary conditions should be satisfied to
infer valuable statistical results:\\
i) The length of measured fluctuations must be large enough. \\
ii) The contribution of superimposed trends and nonstationarities on the
recorded data must be small enough in comparison to intrinsic fluctuations or at least
 distinguishable.\\
Unfortunately in many cases of practical measurements, above
necessary conditions cannot be specified. Identifying trends and
foundation of proper detrending operations are important step toward
robust analysis. Meanwhile, unfortunately, there is no unique
definition of trend and any proper method for extracting it from
underlying data sets in the presence of nonstationarities
\cite{qian10,lim05,hu01,zhi02,golu96,cool65}. On the other hands,
trend in a real world data series especially for non-stationary one,
is an intrinsic function imposed by the nature on data set
\cite{wu07}. To recognize the trend on a data set, one can
investigate the series in whole domain or on some specific span of
domains. Singular value decomposition (SVD) as a filtering procedure
has also been introduced to detrend of signals
\cite{rad05,rad055,sadegh10}. The other method for detrending is
so-called Fourier Detrended Fluctuation Analysis (F-DFA). This
method behaves like a high pass filter and useful for removing
expected sinusoidal trends embedded in underlying data set.

In our previous study concerning plasma fluctuations
\cite{kimiadfa}, we found that the plasma fluctuations in the
various discharge current intensities, behave as anti-correlated
signals. For highly ionized fluid each large deviation from the
electrostatic equilibrium is shielded by a cloud of
oppositely-charged particles \cite{mich94,fran74,fer76,Nishikawa99}.
This also may be related to the fast dissipation of turbulent
kinetic energy in plasma \cite{budev04}.

In this paper, we would like to extend our previous analysis and
open a new insight to reconstruct the stochastic fluctuations and
examine the turbulent feature of underlying data sets. The study of
the universality, non-Gaussianity, multifractality and scaling
behavior of structure function also will be other aims in this
study. At first we should remove trends due to electronic
instrumental systematic noises, alternative current oscillation and
the fluctuations of striation areas near the anode and cathode
plates. As discussed in details in \cite{kimiadfa}, we compute power
spectrum of series using the method proposed in \cite{cool65}. Now
it is easy to track the influence of dominant sinusoidal trends
represented as peaks in the spectral density versus frequency. By
removing these coefficients, actually we diminish their
contributions in the reconstruction of fluctuations. If this part
has been done well, the crossovers in the fluctuations function
given by multifractal detrended fluctuations analysis for
reconstructed series will be disappeared and finally the cleaned
data sets for further investigations will be retrieved. The minimum
number of coefficients in the Fourier space should be eliminated to
remove dominant sinusoidal trends is approximately $400$. It is
worth noting that to find cross-correlation between two different
series Singular Value Decomposition could be used to guarantee the
synchronization. Here using the Markovian method, we explore the
statistical properties of detrended discharge current fluctuations.
Then a Fokker-Planck evolution operator and Langevin equation will
be found \cite{Jafari03,kantz}. To complete our analysis we will
investigate the multifractal exponent derived by Markovian method
and compare it with that of computed by extended self similarity
method. In addition, we discuss about the different sources of the
multifractality in discharge current fluctuations.

The rest of this paper is organized as follows: In Section 2, we
give an explanation of our experimental set up used for recording
plasma fluctuations. Section 3 is devoted to a brief summary of the
most important notions and theorems on Markovian method and their
application to the analysis of empirical data. Using the likelihood
statistics, we determined Markov time scale. Section 4 contains the
main results of our analysis and estimate the Fokker-Planck and
Langevin equations which govern the probability density function and
stochastic variable (discharge current), respectively. An exact
decomposition equation for computing temporal correlation function
and distinguishing between Markov and correlation time scale are
explained in detail in section 4. Non-Gaussianity and
multifractality nature of original and reconstructed time series are
also given in section 5. Section 6 closes with a discussion and
conclusion of the present results.

\begin{figure}
\includegraphics[width=1.0\textwidth]{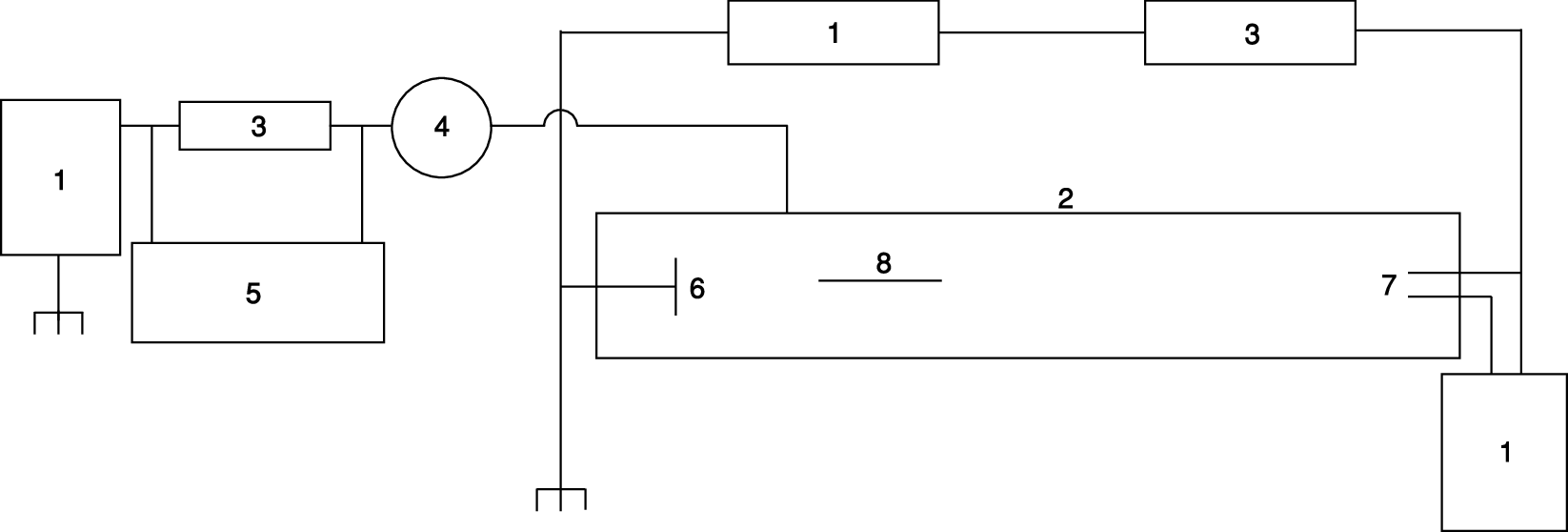}
\caption{The sketch of the experimental
setup used to record the discharge current fluctuations in the tube filled with
Helium, with (1) Power supply; (2) Glass tube; (3) Resistance; (4) Ampere meter; (5) A/D card and PC; (6) Anode plate; (7) Hot
cathode; (8) Single Langmuir probe.}
\label{setup}     
\end{figure}

\begin{figure}
\includegraphics[width=1.0\textwidth]{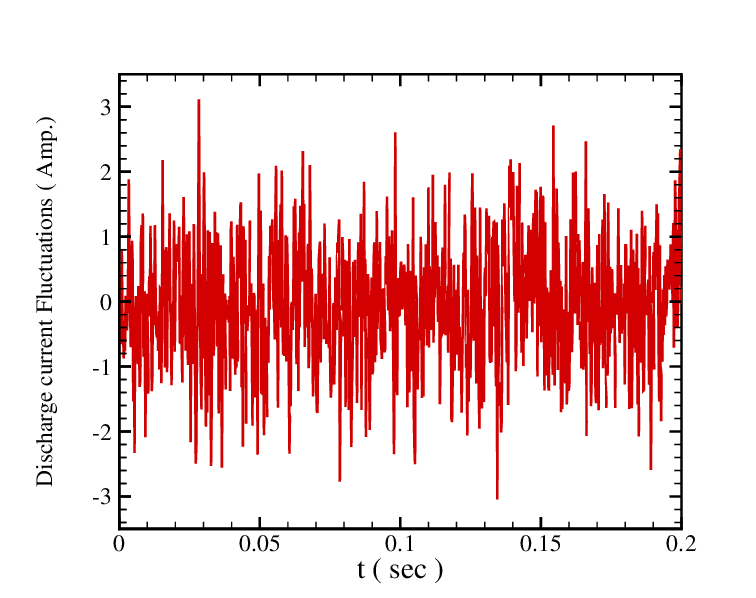}
\caption{Typical
detrended electrical discharge current fluctuations in the plasma as
a function of time}
\label{fig1}       
\end{figure}

\section{Experimental Equipment}
To explore the complex nature of the discharge current fluctuations
in a typical plasma, we constructed an experimental setup as
indicated in Figure \ref{setup}. The plasma chamber has two copper
electrodes attached to the ends of discharge glass tube, 80 mm in
diameter and 110 cm in length. One of these electrode is the anode
(a flat copper plate as a positive pole), while the other one
represents the cathode (as a negative pole and electron propagator).
The base pressure is 0.1 up to 0.8 Torr and the discharge tube is
filled with Helium as the working gas under voltage of $400-900$ V.
The pressure, voltage and current should be fine tuned for ensuring
the stability of the plasma. Using a resistor which was connected to
an operational amplifier impedance converter, discharge current
fluctuations are monitored. In this setup, we fixed the pressure and
examined how the statistical properties of plasma change for various
values of current. The fluctuations of the discharge current
measured by single Langmuir probe were digitized and cleaned with a
filter that omitted direct current. Finally, the fluctuation of the
discharge were recorded by using an analog to digital card for
several values of the electrical discharge current intensity,
namely, 50, 60, 100, 120, 140, 180, and 210 mA at frequency equal to
44100 Hz. The resolution of recorded data is 12 bits. The typical
size of the recorded data sets for each current intensity is about
$10^6$. Cleaned data were constructed by applying the
Fourier-Detrended Fluctuation analysis. Figure \ref{fig1} shows
typical detrended discharge current fluctuation.

\section{Markovian nature of data set}

As mentioned in the introduction, we use the Markovian method to
explore the stochasticity nature of discharge current fluctuations
in plasma. To investigate the Markovian nature of data, we briefly
summarize the conceptions and theorems which will be importance for
our statistical analysis of cleaned data set. For further details on
Markov processes we refer the reader to the references
\cite{fri00,fri97b,Risken,han82,JFM,fri98}.

We represent the discharge current fluctuations as a function of time
by $X(t)$ and define $x(t)=X(t)/\sigma$, where $\sigma$ is the
standard deviation of discharge current fluctuations. Fundamental
quantities related to the Markov processes are conditional
probability density functions. The conditional probability density
function (CPDF), $p(x_2,t_2|x_1, t_1)$, is defined as
\begin{equation}
p(x_2, t_2|x_1, t_1) = \frac{p(x_2, t_2; x_1, t_1)}{ p(x_1, t_1)}
\end{equation}
where $p(x_2, t_2; x_1, t_1)$ is the joint probability density
function (JPDF), describing the probability of finding
simultaneously, $x_1$ at scale(time), $t_1$,  and $x_2$ at
scale(time), $t_2$. Higher order conditional probability densities
can be defined in an analogous way
\begin{equation}
p(x_N, t_N|x_{N-1},t_{N-1}; ...;x_1,t_1) = \frac{p(x_N, t_N;...;
x_1,t_1)} {p(x_{N-1}, t_{N-1}; ...;x_1,t_1)}
\end{equation}
where $p(x_N, t_N; x_{N-1}, t_{N-1}; ...; x_1, t_1)$ is $N$-point
joint probability density function. Intuitively, the physical
interpretation of a Markov process is that it "forgets its past,"
or, in other words, only the most nearby conditioning, namely
$x_{N-1}$ at $t_{N-1}$, is relevant to the probability of finding a
fluctuation $x_N$ at $t_N$. Hence, in the Markov process the ability
to predict the value of $x_N$ will not be enhanced by knowing its
values in the steps prior to the most recent one. So an important
simplification that is made for a Markov process is that, the
conditional multivariate joint PDF is written in terms of the
products of simple two parameter conditional PDF's \cite{Risken} as
\begin{eqnarray}\label{MRR:EQ2}
&&p(x_N,t_N;x_{N-1},t_{N-1};\cdots;x_2,t_2|x_1,t_1)\cr\nonumber \\
&&=\prod_{i=2}^N p(x_i,t_i|x_{i-1}, t_{i-1})\;
\end{eqnarray}

To investigate whether underlying signal is a Markov process, one
should tests the Eq. (\ref{MRR:EQ2}). But in practice for large
values of $N$, is beyond the current computational capability. For
$N=3$ (three points or events), however, the condition will be
\begin{eqnarray}\label{MRR:EQ3}
p(x_3,t_3|x_2,t_2;x_1,t_1)=p(x_3,t_3|x_2,t_2)\;
\end{eqnarray}
which should hold for any value of $t_2$ in the interval
$t_1<t_2<t_3$. A process is then Markovian if the Eq.
(\ref{MRR:EQ3}) is satisfied for a {\it certain} time separation
$t_3-t_2$, in which case, we define the Markov time scale as $t_{\rm
Markov}=t_3-t_2$. For simplicity, we let $t_2-t_1=t_3-t_2$. Thus, to
compute the $t_{\rm Markov}$ we use a fundamental theory of
probability according to which we write any three-point PDF in terms
of the conditional probability functions as
\begin{eqnarray}\label{MRR:EQ4}
&&p(x_3,t_3;x_2,t_2;x_1,t_1)\cr \nonumber
\\ &&=p(x_3,t_3|x_2,t_2; x_1, t_1)p(x_2, t_2; x_1, t_1)
\end{eqnarray}
Using the properties of Markov processes to substitute Eq.
(\ref{MRR:EQ4}), we obtain
\begin{eqnarray}\label{MRR:EQ5}
&& p_{\rm Mar}(x_3,t_3;x_2,t_2;x_1,t_1)\cr \nonumber \\
&& =p(x_3,t_3|x_2,t_2)p(x_2,t_2;x_1,t_1)
\end{eqnarray}
\begin{figure}

\includegraphics[width=1.0\textwidth]{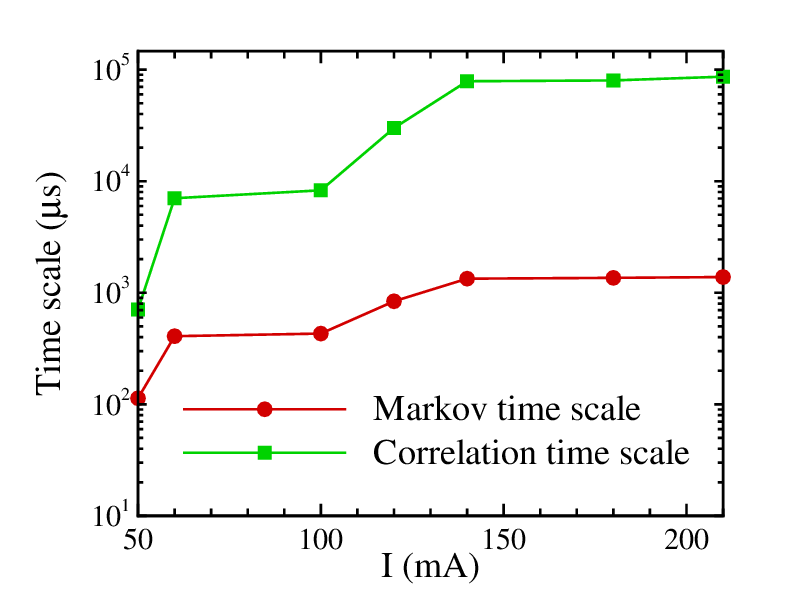}
\caption{Markov and correlation time scales  as a function of
discharge current intensity. The unit of vertical axis is
microsecond.} \label{ff}
\end{figure}
\begin{figure}

\includegraphics[width=1.0\textwidth]{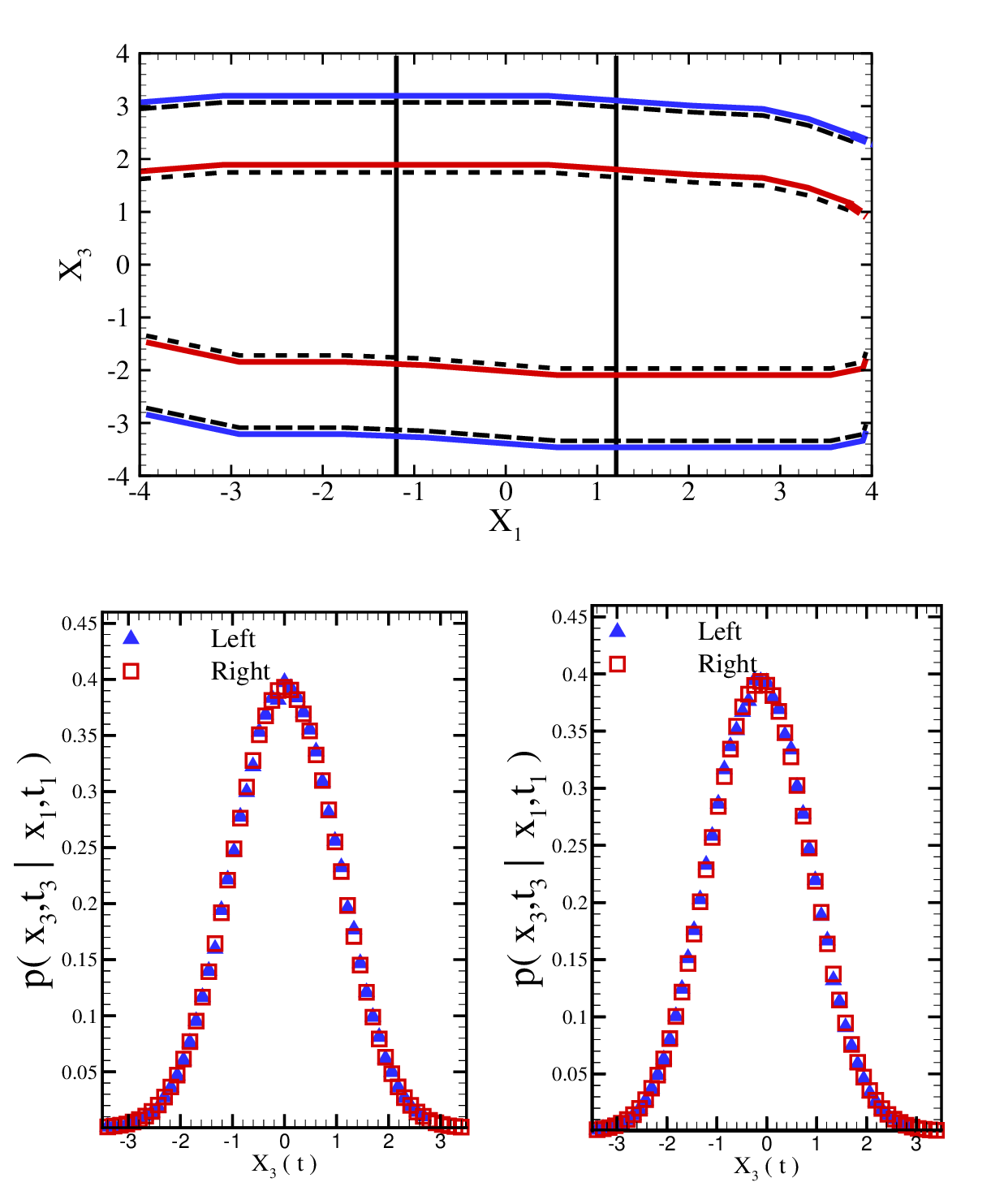}
\caption{Upper panel shows contour plots of the conditional PDF,
$p(x_{3},t_{3}|x_{1},t_{1})$. The solid and dashed line correspond
to the left and right hand side of Eq. (\ref{ck}) for
$t_3-t_1=2\times t_{\rm Markov}$, respectively. Inner contours are a
cutting of PDF at $0.08$ level and outer contours correspond to
$0.005$ level. Lower panel corresponds to the cuts through the
conditional PDF for $x_{1} = \pm 1.25\sigma$.} \label{f50}
\end{figure}

To determine the Markov time scale by means of joint probability
density function (Eqs. (\ref{MRR:EQ4}) and (\ref{MRR:EQ5})), we use
Bayesian statistics  \cite{co04}. We introduce measurements and
model parameters as $\{{\mathcal{X}}\}: \{
p(x_3,t_3;x_2,t_2;x_1,t_1) \}$ and $\{ \Theta \} : \{ t_{\rm
{Markov}}\}$, respectively. Based on the Bayesian theorem, the
conditional probability of the model parameters given data set
(observation) is so-called posterior probability and is given by:
\begin{equation}
P(t_{\rm {Markov}}|{\mathcal{X}})=\frac{{\mathcal{L}}({\mathcal{X}}|t_{\rm {Markov}})P(t_{\rm {Markov}})}{\int
{\mathcal{L}}({\mathcal{X}}|t_{\rm {Markov}})P(t_{\rm {Markov}})d t_{\rm {Markov}}}
\end{equation}
here ${\mathcal{L}}({\mathcal{X}}|t_{\rm {Markov}})$ is the so-called
Likelihood and $P(t_{\rm {Markov}})$ contains all initial constraints
regarding to model parameters, so-called prior distribution expressed the degree of belief about the model. If we have no
any extra information for model free parameters, the posterior function, $P (t_{\rm {Markov}}|{\mathcal{X}})$
is proportional to the Likelihood function. Usually one can consider the various measurements to be independent of each
other, so according to
the central limit theorem, Likelihood function reads as:
\begin{equation}
{\mathcal{L}}({\mathcal{X}}|t_{\rm {Markov}})\sim\exp\left(\frac{-\chi^2(t_{\rm {Markov}})}{2}\right)
\end{equation}
 where:
\begin{eqnarray}\label{chi}
&&\chi^2(t_{\rm {Markov}})=\int dx_1 dx_2 dx_3[p(x_3,t_3;x_2, t_2; x_1,
t_1)\nonumber\\&&-p_{\rm Mar}(x_3,t_3;x_2, t_2; x_1, t_1)]^2/
\left[\sigma_{3-{\rm joint}}^2+\sigma_{\rm Mar}^2\right]
\end{eqnarray}
$\sigma^2_{3-{\rm joint}}$ and $\sigma^2_{\rm Mar}$ are the
variances of $p(x_3,t_3;x_2, t_2; x_1, t_1)$ and $p_{\rm
Mar}(x_3,t_3;x_2, t_2; x_1, t_1)$, respectively. Evidently, when,
for a set of values of the parameters, the $\chi^2(t_{\rm {Markov}})$ is minimized, the
probability will be maximized. The minimum value of $\chi^2_{\nu}(t_{\rm {Markov}})$
($\chi^2_{\nu}(t_{\rm {Markov}}) = \chi^2(t_{\rm {Markov}})/{\cal{N}}$, with ${\cal{N}}$ being the
number of degree of freedom) corresponds to the best value of  $t_{\rm Markov}$ for
different value of electrical discharge current intensities.

The value of error-bar at
$1\sigma$ confidence interval of $t_{\rm {Markov}}$ for each current intensity is determined by the
Likelihood function according to:
\begin{equation}
68.3\%=\int_{-\sigma^{-}}^{+\sigma^{+}}{\mathcal{L}}({\mathcal{X}}|t_{\rm {Markov}})dt_{\rm {Markov}}
\end{equation}

The values of Markov time scales, $t_{\rm Markov}$ in terms of
discharge current intensity have been plotted in Figure \ref{ff}. It
must be pointed out that, the unit of $t_{\rm Markov}$ reported in
this figure has been changed to the units of microsecond ($\mu s$)
by using the rate of digitalization in the experimental setup,
$44100$ $sample/sec$.

One can write Eq. (\ref{MRR:EQ5}) as an integral equation, which is
well-known as the Chapman-Kolmogorov (CK) equation
\begin{equation}\label{ck}
p(x_3,t_3| x_1, t_1)=\int dx_2\;p(x_3,t_3| x_2,t_2)\;p (x_2,t_2|
x_1,x_1)\  \label{ck}
\end{equation}

We have checked the validity of the CK equation for describing the
time scale separation of $t_1$ and $t_2 $ being equal to the Markov
time scale. This is shown in Figure \ref{f50} (for the data set with
electrical current intensity, $I=50$ mA). In this figure, the upper
panel shows the contour plot of identification of the left (solid
line) and right (dashed line) sides of Eq. (\ref{ck}) for two
levels, 0.080 (inner contour) and 0.005 (outer contour). The
conditional PDF $p(x_3,t_3|x_1,t_1)$, for $x_1=\pm 1.25\sigma$, are
shown in the lower panel. All the scales are measured in unit of the
standard deviation of the discharge current fluctuations. We must
point out that if all situations to be same as our experimental
setup such as pressure, current intensity and so on, one can expect
that all values derived by Markov analysis would be repeated. The
value of Markov time scale increases as discharge current intensity
increases (see Figure \ref{ff}). It seems that by increasing the
current intensity, charges become more energetic, therefore their
effective cross-section will decrease and hence increasing their
memory.

Up to now we determined the Markov time scale for each cleaned data
set over which time series behaves as a Markov process. In the next
section we will turn to the deriving master and stochastic equations
governing the evolution of probability density function and
fluctuation itself, respectively.

\begin{table}
\caption{\label{coe1}The values of Kramers-Moyal coefficients for
data set at different discharge current intensities.}
\begin{tabular}{lll}
\hline\noalign{\smallskip}
   &  $D^{(1)}(x)$ & $D^{(2)}(x)$ \\ 
   \noalign{\smallskip}\hline\noalign{\smallskip}
       $50{\rm mA}$ &$-0.160\;x$ &$0.090+0.003\;x+0.070\;x^2$     \\
     $60{\rm mA}$ &$-0.058\;x$ &$0.026+0.002\;x+0.030\;x^2$     \\
          $100{\rm mA}$ &$-0.052\;x$ &$0.026+0.002\;x+0.026\;x^2$     \\
               $120{\rm mA}$ &$-0.028\;x$ &$0.013+0.001\;x+0.014\;x^2$     \\
                    $140{\rm mA}$ &$-0.017\;x$ &$0.008+0.001\;x+0.009\;x^2$     \\
                              $180{\rm mA}$ &$-0.017\;x$ &$0.008+0.001\;x+0.009\;x^2$     \\
 $210{\rm mA}$ &$-0.016\;x$ &$0.009+0.001\;x+0.008\;x^2$     \\
\noalign{\smallskip}\hline
\end{tabular}
\end{table}


\section{The Langevin Equation: Evolution equation to describe the Plasma discharge current fluctuations}

The Markovian nature of the plasma electrical discharge fluctuations
enables us to derive a Fokker-Planck equation - a truncated
Kramers-Moyal equation - for the evolution of the PDF $p(x,t)$, in
terms of time $t$. The Chapman-Kolmogorov (CK) equation, formulated
in differential form, yields the following Kramers-Moyal (KM)
expansion \cite{Risken}

\begin{equation}\label{fokker1}
\label{f22} \frac{\partial}{\partial t} p(x,t)=\sum_{n=1}^{\infty}
\left(- \frac{\partial}{\partial x}\right)^n [D^{(n)}(x,t) p(x,t)]
\end{equation}
where $D^{(n)}(x,t)$ are called as the Kramers-Moyal's coefficients.
These coefficients can be estimated directly from the moments,
$M^{(n)}$, and the conditional probability distributions as
\begin{eqnarray}\label{km}
&& D^{(n)}(x,t)=\frac{1}{n!}\hskip .2cm \lim_{\Delta t \to 0}M^{(n)}
\end{eqnarray}
\begin{eqnarray}
&& M^{(n)}=\frac{1}{\Delta t}\int dx'(x'-x)^n p(x',t+\Delta t|x, t)
\label{d12}
\end{eqnarray}

\begin{figure}
\includegraphics[width=1.5\textwidth]{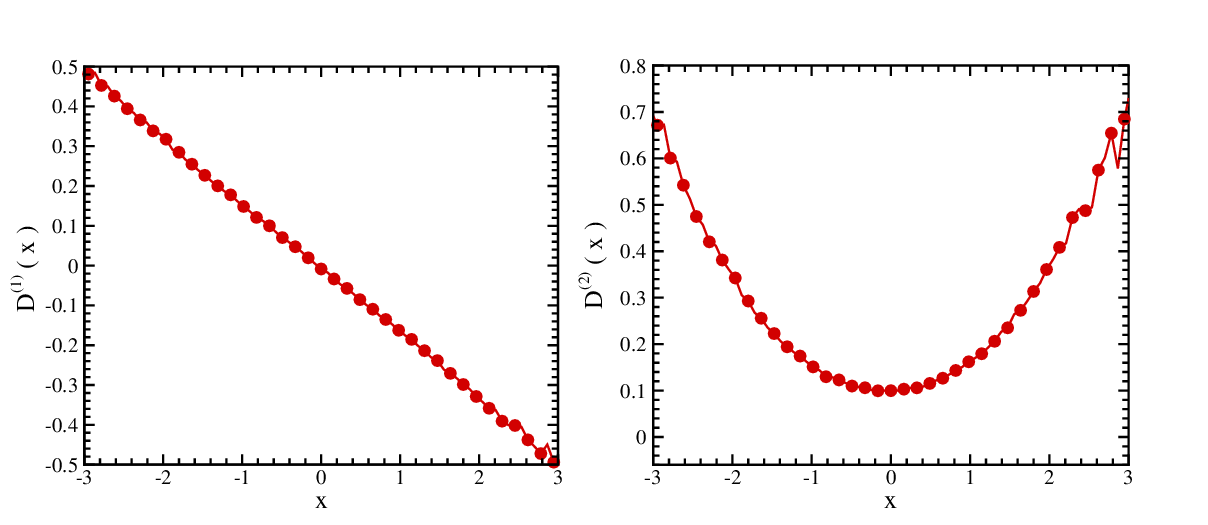}
\caption{Drift,
$D^{(1)}(x)$, diffusion and $D^{(2)}(x)$ coefficients for data set
with $I=50$mA.}
\label{dd}       
\end{figure}

For a general stochastic process, all Kramers-Moyal's coefficients
are different from zero. According to the Pawula's theorem, however, the
Kramers-Moyal expansion stops after the second term, provided that
the fourth order coefficient $D^{(4)}(x,t)$ vanishes.  In that case,
the Kramers-Moyal expansion reduces to a Fokker-Planck equation
(also known as the backwards or second Kolmogorov
equation)\cite{Risken}
\begin{equation}
        \frac{\partial}{\partial t} \, p(x,t) \;=\;
        \left\{ \, - \frac{\partial}{\partial x} D^{(1)}(x,t) \, + \,
        \frac{\partial^2}{\partial x^2} D^{(2)}(x,t) \, \right\}
        p(x,t)\label{foplacond}
\end{equation}
Also the evolution equation for conditional probability density
function is given by the above equation except that $p(x,t)$ is
replaced by $p(x,t|x_1,t_1)$. Here $D^{(1)}$ is known as the drift
term and $D^{(2)}$ as diffusion term which represents the stochastic
part. The Fokker-Planck equation describes the evolution of
probability density function of a stochastic process generated by
the Langevin equation (we use the It\^{o}'s definition)
\cite{Risken}
\begin{equation} \label{Langevin}
 \frac{\partial }{\partial t}x(t)=D^{(1)}(x,t) +
    \sqrt{D^{(2)}(x,t)} f(t)
\end{equation}
where $f(t)$ is a random force, i.e. $\delta$-correlated white noise
in $t$ with zero mean and Gaussian distribution,  $\langle
f(t)f(t')\rangle=2\delta(t-t')$. Using Eqs. (\ref{km}) and
(\ref{d12}), for collected data sets, we calculate  drift, $
D^{(1)}$, and diffusion, $D^{(2)}$, coefficients, shown in Figure
\ref{dd}. It turns out that the drift coefficient $D^{(1)}$ is a
linear function in $x$, whereas the diffusion coefficient $D^{(2)}$
is a quadratic function. For large values of $x$, our estimations
become poor,  the uncertainty increases, so we truncate our
estimations up to $3\sigma$ of fluctuations as indicated in Figure
\ref{dd}.


The functional feature of drift and diffusion coefficients for
different electrical discharge data sets are reported in Table
\ref{coe1}. To ensure that Kramers-Moyal expansion (Eq.
(\ref{fokker1})) reduces to a Fokker-Planck equation (Eq.
(\ref{foplacond})), we compute fourth-order coefficient $D^{(4)}$.
In our analysis, $ {D^{(4)}} \simeq 10^{-1} {D^{(2)}}$. One must
point out that, however the fourth-order Kramers-Moyal's coefficient
is not so small, but in the current analysis, this doesn't make
measurable uncertainty in our results (see below). Furthermore,
using Eq. (\ref{Langevin}), it becomes clear that we are able to
separate the deterministic and the noisy components of the
fluctuations in terms of the coefficients $D^{(1)}$ and $D^{(2)}$.
According to the values of the Kramers-Moyal's coefficients reported
in Table \ref{coe1}, it is possible to reconstruct discharge current
fluctuations at arbitrary current intensity using Eqs.
(\ref{foplacond}) and (\ref{Langevin}) \cite{Jafari03}.

Now let us have a comparison of the statistical properties of
reconstructed data using Eq. (\ref{Langevin}) with the original
fluctuations. For this purpose, we rely on the solution of
Fokker-Planck equation for conditional probability function (same as
Eq. (\ref{foplacond}) for infinitesimally small step $\tau$) which
is given by \cite{Risken}
\begin{eqnarray}
\;p(x_2,t+\tau|x_1,t)&&=\frac{1}{2\sqrt{\pi
D^{(2)}(x_2,t)\tau}}\nonumber\\
&&\times
\exp{\left(-\frac{(x_2-x_1-D^{(1)}(x_2,t)\tau)^2}{4D^{(2)}(x_2,t)\tau}\right)}\label{pjoint1}
\end{eqnarray}

\begin{figure}
\includegraphics[width=1.5\textwidth]{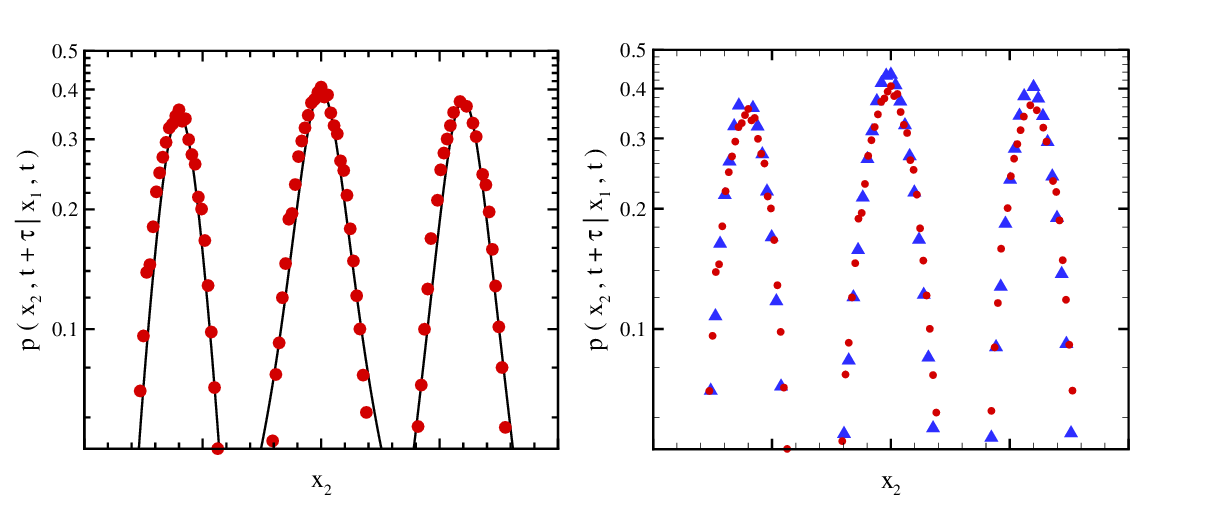}
\caption{Left panel corresponds to the conditional probability
density function determined by analytical formula, Eq.
(\ref{pjoint1}) (solid line) and directly computed by original
cleaned data (symbol) for $I=50$mA. Right panel shows the comparison
between the conditional probability density function determined by
generated data using Eq. (\ref{Langevin}) (triangle symbol) and our
initial cleaned data (circle symbol). In each panel, the plots from
left to right correspond to the cut for $x_1=-0.5\sigma$, $x_1=0.0$
and $x_1=+0.5\sigma$ level, respectively. To make more obvious, we
shifted the value of $x_2$ for each plot. We took $\tau=t_{\rm
Markov}$, where $t_{\rm Markov}$ is the Markov time scale of data
set.} \label{pjoint3}
\end{figure}

Left panel of Figure \ref{pjoint3} shows conditional probability
density function computed by the above equation and directly
calculated from the original detrended data set for $I=50$mA. The
plot from left to right correspond to $x_1=-0.5\sigma$, $x_1=0.0$
and $x_1=+0.5\sigma$ level, respectively. We also compute the
conditional probability using reconstructed fluctuations via Eq.
(\ref{Langevin}) and compare it with the same one for original
cleaned data at three mentioned levels for $x_1$. We took
$\tau=t_{\rm Markov}$ for all plots in Figure \ref{pjoint3}. According to Eq. (\ref{pjoint1}) and based on
Figure \ref{pjoint3} we find a good agreement between stochastic model for reconstructed plasma fluctuations
and original fluctuations.

\begin{figure}
\includegraphics[width=1.5\textwidth]{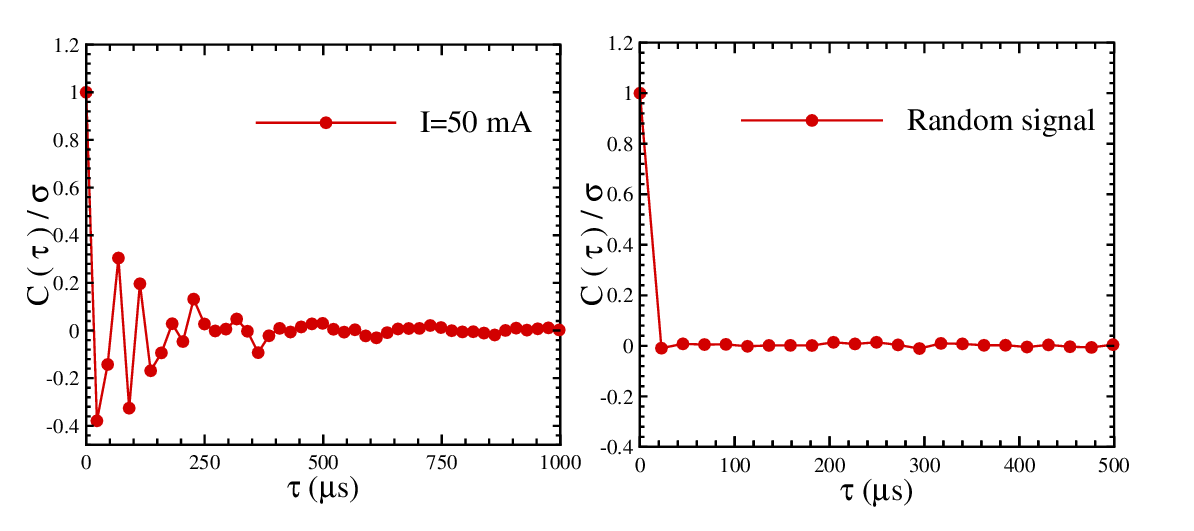}
\caption{Normalized temporal correlation function for plasma cleaned
data for discharge current with $I=50$mA (left panel) and for
completely random data (right panel).}
\label{correlation}       
\end{figure}

According to the definition of Markov time scale, there is no
systematic relation between Markov and autocorrelation time
scales, however one can decompose the temporal correlation
function of an arbitrary stationary Markov processes according to
the formalism introduced by Medvedev \cite{medvedev77}. To this
end, we introduce temporal correlation function as:
\begin{eqnarray}\label{correlation1}
{\mathcal C}(x_1(t_1),x_2(t_2))=\langle x_1(t_1)x_2(t_2)\rangle
\end{eqnarray}
here the sign  $\langle . \rangle$, shows the ensemble averaging. We
have set the mean of time series equal to zero. For a stationary
time series the correlation function depends on only the separation
time scale, which means that
\begin{eqnarray}\label{correlation2}
{\mathcal C}(\tau)&\equiv& {\mathcal C}(x_1(t_1),x_2(t_2))\nonumber\\
&=&\langle x_1(t_1)x_2(t_1+\tau)\rangle
\end{eqnarray}
In the presence of any trends and nonstationarity, correlation
function depends not only to the time separation ($t_2-t_1$), but
also to the starting and finishing times, namely $t_1$ and $t_2$,
respectively. As demonstrated in Ref \cite{kimiadfa}, the underlying
detrended data for discharge current behaves as a stationary signal.
The temporal correlation function for plasma detrended data with
$I=50$mA is plotted in the left hand side of the Figure
\ref{correlation}. This figure confirms the underlying data sets
behave as an anti-correlated series which has been confirmed in
\cite{kimiadfa} using another method. We also present the same plot
for a pure random white noise data, i.e. with Hurst exponent
$H=0.5$, in the right hand side of the Figure \ref{correlation} for
comparison. As explained before the evolution of a typical
stationary Markov process is governed by Master equation (Eq.
(\ref{fokker1})), so the temporal correlation function of this
process can be written as:
\begin{eqnarray}\label{corr1}
\langle x(t+\tau)x(t)\rangle
&=&\sum_{m=0}^{\infty}\frac{|\tau|^m}{m\!}\langle x {\mathcal{F}}^m
x \rangle
\end{eqnarray}
here ${\mathcal{F}}=\sum_{n=1}^{\infty}\frac{D^{(n)}(x,t)}{n\!}\frac{\partial^n}{\partial x^n}$. To calculate temporal correlation function of a Markov process, we should compute probability density of data set. The solution of Eq. (\ref{foplacond}) is:
\begin{eqnarray}
p(x,t)= \frac{const.}{\sqrt{D^{(2)}(x,t)}}\exp\left
(-\int\frac{D^{(1)}(x,t)}{D^{(2)}(x,t)}dx\right )
\end{eqnarray}
If one use the following parameterizations for $D^{(1)}(x,t)=ax(t)$
and $D^{(2)}(x,t)=b+cx(t)+dx^2(t)$, then can find:
\begin{eqnarray}\label{prob2}
p(x,t)=\frac{const.}{\sqrt{D^{(2)}(x,t)}} \exp \left (
\frac{a\ln[b+cx(t)+dx^2(t)]}{2d}-\frac{ac
\arctan{\left ( \frac{c+2dx(t)}{\sqrt{4bd-c^2}} \right )}}{d\sqrt{4bd-c^2}}\right)\nonumber\\
\end{eqnarray}
The constant coefficient could be determined by normalization
procedure. Obviously for $c, d=0$ above probability density function
behaves as a Gaussian distribution. According to the Kramers-Moyal
coefficients reported in Table \ref{coe1} for the plasma
fluctuations, we expect the probability density functions for
various discharge current intensity deviate from exact Gaussian
function. As an example, one can simply show that the temporal
correlation function of a stochastic variable, $x(t)$, governed by
the Langevin equation $\dot{x}(t)=-\gamma x(t)+\eta(t)$ behaves as:
\begin{equation}
\langle x(t+\tau)x(t)\rangle \sim e^{-\gamma \tau}
\end{equation}
consequently one can introduce temporal correlation scale as
$\gamma^{-1}$ while for this process, Markov length scale equates to
unity \cite{medvedev77}. This shows that for our data set the
correlation time scale is greater or equal to the Markov time scale.
In Figure \ref{ff}, we estimate the temporal correlation scale at
stationary case of plasma fluctuations by using $D^{1}(x,t)$
reported in Table \ref{coe1}. In addition for a scaling behavior of
autocorrelation function, namely $\langle x(t+\tau)x(t)\rangle\sim
\tau^{-\kappa}$, in terms of Hurst exponent, one can find out the
scaling exponent for autocorrelation function as $\kappa=2-2H$. Then
by increasing Hurst exponent, $\kappa$ decreased and degree of
correlation to be increased (see e.g. \cite{kimiadfa}). Our results
show that Markov time scale is almost an increasing function versus
discharge current intensity which directly reflects the memory in
the stochastic current fluctuations produced in plasma measured by
Langmuir probe.


\section{Non-Gaussianity and Multifractality of Plasma fluctuations }

 In this section we investigate the Gaussian
nature of the PDFs of reconstructed and detrended time series as
well as its multifractal exponent derived by Markovian approach. For
a Gaussian distribution, all the even moments are related to the
second moment
 through $\langle x^{2n}\rangle=\frac{2n!}{2^nn!}\langle
x^2\rangle^n$ (e.g., for $n=2$, $\langle x^{4}\rangle=3\langle
x^2\rangle^2$), while the odd moments are zero identically. We can
directly check the relation between the higher moments for the
plasma fluctuations data at different value of discharge current
intensities with second moment. The values of moments and their
variances calculating directly from data are summarized in Table
\ref{moments}.


\begin{table}
\caption{\label{moments}The values of moments, $\langle x^n\rangle$,
and their errors for data set at different discharge current
intensities.}
\begin{tabular}{llll}
\hline\noalign{\smallskip}
   &  $\langle x^2\rangle\times 10^{+5}$ & $\langle x^3\rangle\times 10^{+9}$&$\langle x^4\rangle\times 10^{+9}$   \\
\noalign{\smallskip}\hline\noalign{\smallskip}
   &&&\\
       $50{\rm mA}$ &$2.483\pm 0.001 $ &$-0.395\pm0.417$&$1.842\pm 0.062$\\
     &&&\\$60{\rm mA}$ &$2.415\pm 0.001$ &$-3.330\pm 0.123$&$1.532\pm 0.001$
     \\
          &&&\\$100{\rm mA}$ &$2.879\pm 0.001$ &$-17.500\pm 1.990$&$4.140\pm 0.318$
     \\
               &&&\\$120{\rm mA}$ &$2.600\pm 0.001$ &$-3.140\pm 0.179$&$1.858\pm 0.008$
     \\
                    &&&\\
                    $140{\rm mA}$ &$2.095\pm 0.001$ &$-5.770\pm 0.380$&$1.431\pm 0.030$
     \\

                &&&\\ $180{\rm mA}$ &$1.617\pm 0.001$ &$-7.220\pm 2.850$&$2.002\pm0.765$
     \\ 
 &&&\\$210{\rm mA}$ &$1.383\pm 0.001$ &$-0.458\pm 0.139$&$0.811\pm 0.335$\\
\noalign{\smallskip}\hline
\end{tabular}
\end{table}

Let us examine the predictions for the moments of the plasma
fluctuations via the Fokker-Planck equation, and compare their
values with the direct evaluation represented in the Table
\ref{moments}. Using the general Kramers-Moyal expansion, Eq.
(\ref{fokker1}), which is also valid for the probability density
$p(x,t)$, differential equations for the $n$-th order moments can be
derived. By multiplication of the both side of Eq. (\ref{fokker1})
with $x^n$ and integration with respect to $x$, we can obtain
evolution of different moments of data set as:
\begin{eqnarray}
        \frac{d}{dt} \left< x^n(t) \right> &\, =\,&\sum\limits_{k=1}^{\infty} \left( -1 \right)^k\int_{-\infty}^{+\infty} x^n \left( \,
        \frac{\partial}{\partial x}\right)^k D^{(k)}(x,t)p(x,t) dx
        \nonumber \\ &\,=\,& \sum\limits_{k=1}^{n} \, \frac{n!}{(n-k)!} \,
        \int_{-\infty}^{+\infty} \, x^{n-k} \, D^{(k)}(x,t) \,
        p(x,t) dx
         \nonumber \\
        &\,=\,& \sum\limits_{k=1}^{n} \, \frac{n!}{(n-k!)} \, \left< \,
        x^{n-k} \, D^{(k)}(x,t) \, \right>
        \label{momentengleichung}
\end{eqnarray}
We put $n=4$ in the above equation and find the equation for the
fourth moment as follows
\begin{eqnarray}
\label{moment4} &&\frac{d}{dt}\langle x^4(t)\rangle=4\langle
D^{(1)}(x)x^3(t)\rangle+12 \langle
D^{(2)}(x)x^2(t)\rangle\nonumber\\
&&\qquad +24\langle D^{(3)}(x)x(t)\rangle+24\langle
D^{(4)}(x)\rangle
\end{eqnarray}

The third and fourth Kramers-Moyal's coefficients for the data set
are reported in Table \ref{34}. We should point out that the values
of $|D^{(3)}|$ and $|D^{(4)}|$ are less than $|D^{(2)}|$. For the
stationary case, all the moments of fluctuations are time
independent and the left-hand side of Eq. (\ref{moment4}) vanishes,
so
\begin{eqnarray}
\label{ns}\langle x^4\rangle&=&[\alpha_2(I)\pm \sigma_2(I)]\langle
x^2\rangle^2+[\alpha_3(I)\pm\sigma_3(I)]\langle x^3\rangle
\sqrt{\langle x^2\rangle}\nonumber\\
\end{eqnarray}
where $\alpha_2(I)$ determines the coefficient of kurtosis quantity
and $\sigma_2(I)$ shows its variance. Also $\alpha_3(I)$ determines
the coefficient of skewness and $\sigma_3(I)$ indicates its error.
The skewness measures the asymmetry of probability density function
and kurtosis determines the statistic of rare events in the
processes. In generally they may depend to the discharge current
intensity, $I$. Using the results represented in the Table \ref{34},
for each case of fluctuations $\alpha_n(I)$ and its variance are
given in the Table \ref{alpha}.

\begin{figure}
\includegraphics[width=1.\textwidth]{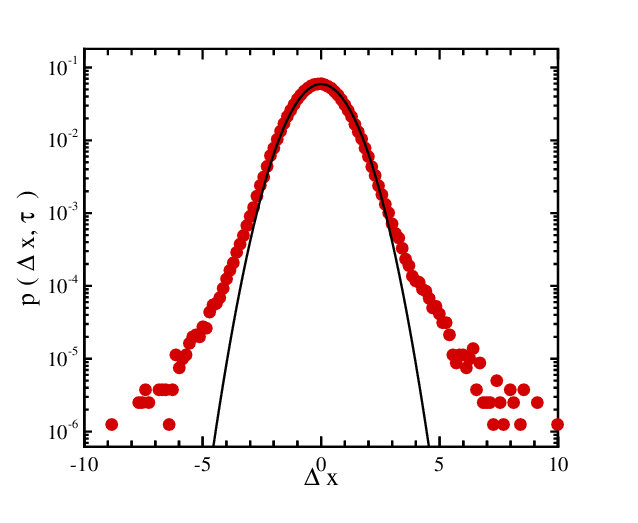}
\caption{Probability density function of reconstructed increment data set (filled symbols) and a typical Gaussian function (solid line) for $I=50$mA and $\tau=1$.}
\label{pdelta}       
\end{figure}

\begin{table}
\caption{\label{34}The values of third and fourth Kramers-Moyal
coefficients for data set at different discharge current
intensities.}
\begin{tabular}{lll}
\hline\noalign{\smallskip}
   &  $D^{(3)}(x)$ & $D^{(4)}(x)$ \\ 
   \noalign{\smallskip}\hline\noalign{\smallskip}

       &&\\$50{\rm mA}$ &$-0.007-0.073\;x$&$0.009+0.009\;x+0.010\;x^2$\\
       &$-0.019\;x^3$&$-0.001\;x^3+0.001\;x^4$\\&&\\
     &&\\$60{\rm mA}$ &$-0.024\;x-0.002\;x^2$&$0.001-0.020\;x+0.010\;x^2$\\
     &$-0.010\;x^3$&$+0.001\;x^3+0.003\;x^4$\\&&\\ 
&&\\$100{\rm mA}$ &$-0.025\;x-0.001\;x^2$&$0.002+0.001\;x+0.013\;x^2$     \\
    &$-0.009\;x^3$&$+0.002\;x^4$\\&&\\  
        &&\\ $120{\rm mA}$ &$-0.013\;x-0.005\;x^3$&$0.001+0.006\;x^2+0.001\;x^4$   \\
         &&\\  
         &&\\ $140{\rm mA}$ &$-0.008\;x-0.003\;x^3$&$0.0009+0.003\;x^2+0.001\;x^4$     \\
          &&\\&&\\  
        &&\\$180{\rm mA}$ &$-0.007\;x-0.003\;x^3$&$0.0007+0.002\;x^2+0.001\;x^4$     \\
        &&\\&&\\  
 &&\\$210{\rm mA}$ &$-0.009\;x-0.002\;x^3$&$0.0008+0.005\;x^2$\\&&
 \\ 
\noalign{\smallskip}\hline
\end{tabular}
\end{table}
\begin{table}
\caption{\label{alpha} The values of $\alpha$'s coefficients and
their variances, $\sigma$'s, for data set at different discharge
current intensities.}
\begin{tabular}{lllll}
\hline\noalign{\smallskip}
   &  $\alpha_2$&$\sigma_2$ & $\alpha_3$&$\sigma_3$ \\ 
\noalign{\smallskip}\hline\noalign{\smallskip}
       $50{\rm mA}$&$3.14$& $0.38$&$0.18$ &$0.21$     \\ 
     $60{\rm mA}$ &$3.81$&$0.70$&$-1.17$ &$0.51$     \\ 
          $100{\rm mA}$&$2.86$& $0.53$&$-3.68$ &$0.66$     \\ 
               $120{\rm mA}$ &$2.77$&$0.98$&$2.25$ &$0.96$     \\ 
                    $140{\rm mA}$ &$3.05$&$0.56$&$-2.09$ &$0.35$     \\ 
                              $180{\rm mA}$&$3.67$&$0.85$ &$-0.01$ &$0.28$     \\
 $210{\rm mA}$&$3.08$&$0.95$ &$-4.18$ &$0.95$     \\ 
\noalign{\smallskip}\hline
\end{tabular}
\end{table}
As we mentioned before, for exact Gaussian process, we should have
\begin{eqnarray}
\alpha_2(I)&=&\frac{\label{ns}\langle x^4\rangle}{\langle
x^2\rangle^2}=3.0\\
\alpha_3(I)&=&0.0
\end{eqnarray}
If $\alpha_2(I)>3.0$ means that probability density function has fat
tail and rare events have more chance to occur (with respect to the
Gaussian process). While for $\alpha_2(I)<3.0$, the tails  of
probability density function is heavy than the Gaussian
distribution. According to the values of $\alpha_2(I)$ and
$\alpha_3(I)$, we find that probability density function of data set
is deviated from Gaussian. This deviation can be characterized by
skewness as well as kurtosis coefficients \cite{dynk84}. Table 4
demonstrates that there is no monotonous behavior for deviations
from Gaussianity as a function of discharge current intensity
\cite{prim,phb05}. Subsequently the properties of probability
density function appears almost independent of the plasma conditions
\cite{sat09}. In this case, we expect that the increment of signals
also may reveal the non-Gaussianity properties. To this end, we
introduce increment series as $\Delta x(\tau)\equiv x(t+\tau)-x(t)$,
where $\tau$ is time delay. We do the same computation to determine
whether this new data set has Markovian nature. Our analysis
demonstrate that Markov time scale of $\Delta x(\tau)$ for all
discharge current intensity is  $136_{-45}^{+90} \mu s$. Figure
\ref{pdelta} shows the probability density of reconstructed
increment data set with a typical time lag equates to $\tau=1$ for
$I=50$mA. If the probability density function to be fatter than
Gaussian function hence the probability of observing fluctuations
far exceeding the average amplitude are not ignorable. This
phenomenon can affect on usual transport in the plasma.

To check the multifractal nature of reconstructed time series, we
investigate the Markovian nature of the increments of profile which is defined
as: $\Delta x(\tau)\equiv y(t+\tau)-y(t)$, where $y(t)=\sum_{i=0}^{t}x(i)$. For convenience, hereafter we rename
$y(t)$ by $x(t)$.  According to the mentioned
procedure, we can determine the Markov time scales for the
increments and calculate the Kramers-Moyal's coefficients.
Likelihood analysis confirms that, the increment of profile signal for all electrical current intensities are also Markov
processes.

The Fokker-Planck equation for probability density
function of the increment is given by \cite{fri00,fried01}
\begin{eqnarray}
&&-\tau\frac{\partial }{\partial \tau}\,p(\Delta x,\tau)
=\nonumber\\&& \left\{- \frac{\partial}{\partial \Delta x}
D^{(1)}(\Delta x,\tau)+\frac{\partial^2}{\partial \Delta x^2}
D^{(2)}(\Delta x,\tau)\right\} p(\Delta x,\tau)
\label{fopincrement1}
\end{eqnarray}
the negative sign of the left-hand side of Eq. (\ref{fopincrement1})
is due to the direction of the cascade from large to smaller time
scales $\tau$. The corresponding Langevin equation can be read as
\begin{equation} \label{Langevinincrement}
-\tau \frac{\partial }{\partial \tau}\Delta x(\tau)=D^{(1)}(\Delta
x,\tau) +
    \sqrt{D^{(2)}(\Delta x,\tau)} f(\tau)
\end{equation}
where $f(\tau)$ is the same as random function in Eq.
\ref{Langevin}. For time series with scaling correlations the drift
and diffusion coefficients of increment are formulated as
\cite{fri00,fried01,surface}
\begin{eqnarray}
D^{(1)}(\Delta x, \tau)&\simeq&-H\Delta x\nonumber\\ D^{(2)}(\Delta
x, \tau)&\simeq&b\Delta x^2 \label{fopincrement2}
\end{eqnarray}

Using Eqs. (\ref{fopincrement1}) and (\ref{fopincrement2}) we obtain
the evolution of structure functions as: (
$S_q(\tau)\equiv\langle|\Delta
x(\tau)|^q\rangle=\langle|x(t+\tau)-x(t)|^q\rangle$) as follows
\begin{eqnarray}
-\tau\frac{\partial}{\partial \tau}\langle|\Delta
x(\tau)|^q\rangle&=& q\langle|\Delta x(\tau)|^{q-1}D^{(1)}(\Delta
x,\tau)\rangle\nonumber\\&&+q(q-1)\langle|\Delta
x(\tau)|^{q-2}D^{(2)}(\Delta x,\tau)\rangle \label{fopincrement3}
\end{eqnarray}
by substituting the Eqs. (\ref{fopincrement2}) in Eq.
(\ref{fopincrement3}) we find
\begin{eqnarray}
\tau\frac{\partial}{\partial \tau}\langle|\Delta
x(\tau)|^q\rangle&=& [qH-bq(q-1)]\langle|\Delta x(\tau)|^{q}\rangle
\label{fopincrement4}
\end{eqnarray}
the above equation implies scaling behavior for moments of
increments, structure function as
\begin{eqnarray}
S_q(\tau)\equiv\langle|\Delta
x(\tau)|^q\rangle=\langle|x(t+\tau)-x(t)|^q\rangle\sim\tau^{\xi(q)}.
\label{momincrement}
\end{eqnarray}

According to Eqs. (\ref{fopincrement4}) and (\ref{momincrement}),
the corresponding scaling exponent in general case can be read as
\begin{eqnarray}
\xi(q)=qH-bq(q-1)\label{xi1}
\end{eqnarray}
For mono- and multi-fractal processes the exponent $\xi(q)$ have
linear and non-linear behavior with $q$, respectively. It must point
out that $H$ is nothing except the underlying fluctuations's Hurst
exponent \cite{hurst65,eke02,koscielny98,koscielny98b}.
The obtained expression for  $D^{(1)}(\Delta x, \tau)$ and
$D^{(2)}(\Delta x, \tau)$ (to avoid the overissue we just report the
results of data for $I=50$mA) are as follows
\begin{eqnarray}
D^{(1)}(\Delta x, \tau)&=&-(0.45\pm0.03)\Delta x\nonumber\\
D^{(2)}(\Delta x, \tau)&=&(0.04\pm0.01)\Delta x^2
\label{fopincrement50}
\end{eqnarray}
consequently, using Eqs. (\ref{fopincrement50}) and (\ref{xi1}), the
scaling exponent is determined as
\begin{eqnarray}
\xi(q)=(0.45\pm0.03)q-(0.04\pm0.01)q(q-1)\label{xi2}
\end{eqnarray}

\begin{figure}
\includegraphics[width=1.5\textwidth]{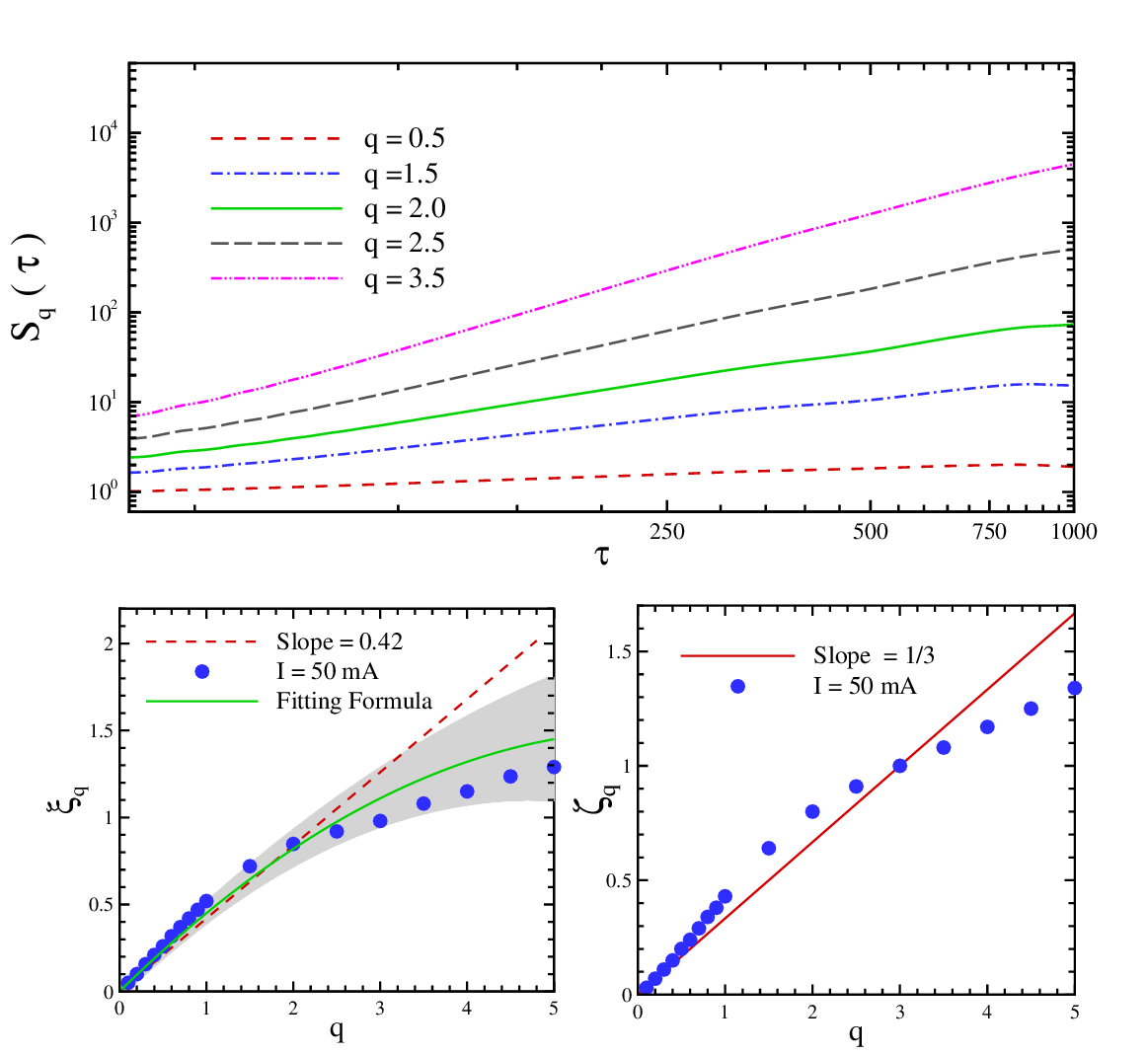}
\caption{Upper
panel indicates the structure function versus $\tau$. Lower left
panel shows the scaling exponent of $S_q(\tau)$ as a function of
moment for original cleaned plasma fluctuations (filled symbol) and
solid line corresponds to the fitting formula derived by
Kramers-Moyal's coefficients for multi-fractal anti-correlated
signal with $H=0.42$ (see Eq. (\ref{xi2})). Also in this panel,
dashed line corresponds to a mono-fractal anti-correlated series.
Lower right panel indicates $\zeta(q)$ versus $q$. Here we chose the
data set with $I=50$mA.}
\label{s2}
\end{figure}

To check the consistency of  estimated scaling exponent $\xi(q)$,
Eq. (\ref{xi2}), with that of determined by original time series we
use the extended self similarity (ESS) method \cite{benzi96,ber03}.
Extended Self Similarity is a method to find an extended range of
scaling behavior of underlying stochastic fluctuations. The
prediction of Kolmogorov (K41) theory for the velocity field of
fully developed turbulence namely in the inertial regime is
$S_q(\tau)\sim \tau^{\xi_q}$ with $\xi_q=\frac{q}{3}$ and shows a
nonofractal behavior \cite{kolm41}. The deviation from this
prediction have been reported experimentally and theoretically, due
to the energy dissipation fluctuations (see.
\cite{ber03,ans84,benzi84,menev87,ess93,ghasemiess}). In the context
of Extended Self Similarity, the self similarity expressed above to
be changed to a new scaling relation according to $S_q\sim
S_3^{\zeta_q}$ in which not only the scaling regime for self
similarity behavior to be extended even further from inertia range,
but also the statistical uncertainty for determining scaling
exponent decreases. In the Extended Self Similarity method, the
log-log plot of $S_q(\tau)$ as a function of specific order of
structure function, namely $S_3(\tau)$, usually shows an extended
scaling regime
\begin{eqnarray}
S_q(\tau)\sim S_3(\tau)^{\zeta (q)}. \label{ess}
\end{eqnarray}
For any Gaussian process, the exponent in the above equation is
given by $\zeta(q)=q/3$ \cite{benzi96,ber03}. Any deviation from
this relation can be interpreted as a deviation from Gaussianity.
Figure \ref{s2} shows the log-log plot of structure function in
terms of time scaling (upper panel), exponents $\xi(q)$ (left lower
panel) and $\zeta(q)$ (right lower panel) for the plasma
fluctuations with $I=50$mA. The present results are in agrement with
our previous results derived that the plasma time series have
multi-fractal nature \cite{kimiadfa}.

As shown in the lower left panel of Figure \ref{s2}, Eq.  (\ref{xi2})
for $\xi(q)$ (solid line) with the shaded area corresponds
to $68.3\%$ confidence interval derived by Markovian analysis of
increment of profile has an acceptable confidence level to experimental
results (filled symbol).

\section{Summary and Conclusion}

Many methods have been devoted to study the fluctuations in the
plasma \cite{gent95,krom02,oberm83,taylor63,hazelt04}. Since,
discharge current fluctuations can serve as a quantitative indicator
of plasma disturbances, consequently any tantalizing statistical
evidences give new insight throughout plasma fluctuations. We have
studied the stochastic nature of the electrical discharge current
fluctuations in the Helium plasma as a working gas. As mentioned
before, fluctuations measured by Langmuir probe assimilate
stochastic phenomena occurred in a typical plasma fluid. Therefore
it can reveal many interesting feature which can not investigated by
common methods in data analyzing. We have applied the
Fourier-Detrended Fluctuations Analysis method to extract sinusoidal
trend and used clean data set for further analysis \cite{kimiadfa}.

Here we used the novel approach i.e. the Markovian method to
investigate many statistical properties of the current fluctuations
in the plasma. We showed that how the mathematical framework of
Markov processes can be applied to develop a successful statistical
description of the plasma fluctuations. We have analyzed detrended
data via Markovian method. The Markov time scale, as the
characteristic time scale of the Markov properties of the electrical
discharge current fluctuations, was obtained. According to the
theory of the stochastic process, the electrical discharge
fluctuations at time scales larger than the Markov time scale can be
considered as a Markov process. This means that the data located at
the separations larger than the Markov time scale can be described
as a Markov chain. It is found that Markov time scale, $t_{\rm
Markov}$ increases by increasing the current intensity in the
plasma. This means that the memory of charged particles in the
plasma increase as current intensity increases. It is due to the
fact that particles become more energetic, therefore they can
penetrate deeper in the plasma without considerable deviation from
the initial trajectory.  In other words, by increasing discharge
current density the electron impact ionization cross section almost
decreases, consequently it is statistically expected that memory of
electrons at this mesoscale for energy transfer to be decreased
causing the drift as well as diffusion coefficients of current
fluctuations to be reduced \cite{anderson96,beush00,cross}.

Using the Markovian nature of fluctuations, we demonstrated that,
the probability density function of fluctuations satisfies a
Fokker-Planck equation. Based on this equation one can do averaging
to extract relevant observable quantities of plasma fluctuations.
The so-called Kramers-Moyal's coefficients by using conditional
moments (Eq. (\ref{d12})) have been determined. The Langevin
equation, governing the evolution of current fluctuations has also
been given. To check the consistency of statistical properties of
regenerated data set with original cleaned series, we compare
conditional probability density function derived by Eq.
(\ref{pjoint1}) with that of computed by original data in Figure
\ref{pjoint3}.

By using exact decomposition of temporal correlation function for
stationary Markov processes, we gave an expression (Eq.
(\ref{corr1})) to determine correlation function of plasma
fluctuations. For the stationary time series, we calculated
correlation time scale which in principle differs from Markov
characteristic time scale. We argued that there is no systematic
relation between Markov and correlation time scales. It must point
out that, Markov time scale is potentially related to energy
transfer in mesoscale dynamics. Also here based on Markovian method
we gave an equation for evolution of various moments of structure
function (Eq. (\ref{momentengleichung})). As we expected from Eq
(\ref{prob2}), a deviation from Gaussianity has been observed. This
might give a hint toward the multifractality nature of plasma
fluctuations in our set up.  Our results confirm that plasma
fluctuations in all range of current intensity prepared in our set
up behave as non-Gaussian processes. To extend the scaling behavior
of $S_q(\tau)$ versus $\tau$, we relied on Extended Self Similarity
method. Extended Self Similarity approach confirmed that, the
scaling exponents of the discharge current fluctuations didn't
follow the Kolmogorov (K41) scaling exponents. It means that there
is no constant energy cascade form large scales (time or space) to
small one and there exists energy dissipation fluctuations in the
plasma. 

\begin{acknowledgements}
 Authors would like to thank  S. Sobhanian
for useful comments and discussions.
\end{acknowledgements}

\end{document}